\documentclass{epl}
\usepackage{amsmath}
\usepackage{amssymb}
\usepackage[dvips]{color}
\usepackage{psfrag}
\usepackage{subfigure}
\usepackage{graphics}
\usepackage{graphicx}
\usepackage{bm}

\newcommand{\be}{\begin{eqnarray}}

\newcommand{\Tet}{T^{*}(L)}

\newcommand{\ff}[2]{\frac{#1}{#2}}
\newcommand{\dd}[0]{\mathrm{d}}
\newcommand{\bb}[0]{\begin{eqnarray}}
\newcommand{\ee}[0]{\end{eqnarray}}
\newcommand{\nn}{\nonumber}
\newcommand{\fd}{\frac{1}{2}}

\newcommand{\PF}[0]{\mathcal{Z}}

\newcommand{\D}{\mathcal{D}}

\newcommand{\action}{\mathcal{S}}

\newcommand{\Tc}{T_\mathrm{c}}

\newcommand{\qv}[0]{{\bm{q}}}
\newcommand{\ttha}{\mathcal{T}}

\title{Origin of the approximate universality of distributions\\ in equilibrium correlated systems}
\author{Maxime Clusel\inst{1}
   \and Jean-Yves Fortin\inst{2} \and Peter C.W. Holdsworth\inst{3}}
\institute{
  \inst{1} Institut Laue-Langevin, 6 rue J. Horowitz, 38042 Grenoble cedex, France\\
 \inst{2} Laboratoire Poncelet, CNRS/UMI 2615, Bolshoy Vlasyevskiy Pereulok 11, Moscow 119002, Russia\\
  \inst{3}Laboratoire de Physique, \'Ecole normale sup{\'e}rieure de Lyon, 46, All{\'e}e d'Italie, 69007 Lyon, France
}
\shortauthor{M. Clusel \etal}
\shorttitle{universality of distributions in equilibrium systems}
\begin{document}
\maketitle
\begin{abstract}
We propose an interpretation of previous experiments and numerical
experiments showing that, for a large class of systems,
distributions of global quantities are similar to a distribution
originally obtained for the magnetization in the 2D-XY model
\cite{BHP}. This approach, developed for the Ising model, is based
on previous numerical observations \cite{CFH1}. We obtain an
effective action using a perturbative method, which successfully
describes the order parameter fluctuations near the phase
transition. This leads to a direct link between the D-dimensional
Ising model and the XY model in the same dimension, which appears
to be a generic feature of many equilibrium critical systems and
which is at the heart of the above observations.\vspace{5mm}\\ 
\textbf{Accepted for publication in Europhysics Letters.}
\end{abstract}
Following a first observation by Bramwell, Holdsworth and Pinton
\cite{BHP}, many studies report that the probability density
functions (PDF), for spatially or temporarily averaged quantities
in correlated equilibrium \cite{bramwell00} and out-of-equilibrium
\cite{LPF96,gleeson,vanmil} systems have a generic asymmetric form
similar to the so-called BHP distribution. This distribution (see
figure \ref{testb}) is obtained from the magnetic fluctuations in
the two-dimensional (2D) XY model of magnetism, in the spin wave
approximation, in the zero temperature limit \cite{bramwell01}.
This "superuniversality" is clearly incompatible with the notion
of universality classes in critical phenomena and its ubiquitous
presence in out-of-equilibrium phenomena appears mysterious to say
the least. In fact it is easy to find critical systems where the
distribution is radically different and in most situations, some
deviation from BHP distribution is apparent. Even in the case of
the 2D-XY model itself it has been recently established that small
temperature-dependent corrections exist
\cite{Palma02,Banks05,Mack05}. There are therefore strict physical
criteria associated with both the observation of this generic
behaviour and with the deviations from it, just as in the case of
Gaussian fluctuations through the application of the central limit
theorem.
 In this paper we expose these criteria
through the study of a well known and well controlled equilibrium
system, the 2D Ising model, the aim being to show microscopically
how a XY-like behaviour appears in the Ising model.\\
Using numerical simulations for the 2D Ising model
\cite{bramwell00,CFH1} we established that there exists a range of
temperatures $T^*(L)$, or applied field $H^*(L)$ close to the
critical point\footnote{$T^*$ was defined in ref.~\cite{CFH1} as
the temperature where the kurtosis of the distribution most
closely approximates to that for the BHP function. As usual for
finite size systems, we define the critical temperature the one
corresponding to the maximum value of the susceptibility.}
$(\Tc(L), H=0)$, where $L$ is the system size, where the PDF is
similar to the BHP function. At this specific temperature or field
the magnetization shows intermittent behaviour, where coherent
structures appear on intermediate time scales, in analogy with
injected power fluctuations in enclosed turbulent
flow~\cite{LPF96,CFH1}. This should be contrasted with the
behaviour at the critical point, where the amplitude of the
structures and the ensuing intermittency are cut off by the
boundaries of the available phase space. The cross over from
approximate superuniversality to universality class dependence is
related to this change of regime. We approach the critical point
from the ordered phase, making a perturbation expansion about the
ordered state. The calculation shows that there is indeed a
quantitative similarity between the fluctuations for 2D XY and
Ising models up to this threshold and this is the origin of
generic behaviour in the Ising system. Through this result we are
able to make some precise statements about the criteria leading to
such generic
behaviour in more disparate and less well controlled systems.\\
We consider the 2D classical Ising model on a square lattice of
size $L\times L$ described, after the Hubbard-Stratronovitch
transformation, by $N=L^2$ continuous variables $\phi_i$, leading
to the partition function $\PF$ \cite{parisi}:
\bb
\PF\propto \left( \ff{\mathrm{det}\bm{K}}{2\pi}\right)^{1/2}\: \int \D \phi \:
\exp\left[-\frac{1}{2}\sum_{i,j=1}^N\phi_iK_{ij}\phi_j+\sum_{i=1}^N\log
\cosh\left(\sum_{j=1}^N K_{ij}\phi_j+\beta H\right)\right],
\label{resHS}
\ee
where $\bm{K}$ is the coupling matrix whose elements are
$K_{ij}=K(2\lambda \delta_{r_i,r_j}+\sum_{\delta}\delta_{r_j,r_i+
\delta} )$, and $\beta=K=1/T$. The vectors $\delta$ are the
lattice unit vectors $\pm \hat x$ and $\pm \hat y$, and $\lambda$
is an arbitrary parameter introduced by the Hubbard-Stratonovitch
transformation. It is chosen so that all the eigenvalues of
$\bm{K}$ are positive, and the mean-field critical temperature
$\Tc^{\mathrm{mf}}=2(D+\lambda)$ is usually defined with
$\lambda=0$. The local spin $\sigma_i$ is thus mapped onto the
local magnetization $m_i=\tanh\left(\sum_{j=1}^N K_{ij} \phi_j +
\beta H \right)$.
\section{Generalized Ginsburg criterion for fluctuations at large length scale}
The usual criterion used in critical phenomena to discuss the
validity of a perturbative approach to the fluctuations is the
Ginsburg criterion. This is defined such that the ratio of the
two-point correlation function, averaged up to the correlation
length $\xi$ and the square of the magnetization, averaged on the
same scale, be small. For the 2D Ising model it is well known that
this ratio is not small compare to unity, and therefore a
perturbative approach can not capture the physical behaviour of
the model up to this length scale. In our particular case however
we are interested in the fluctuations at a large scale $l$, not
equal to $\xi$. It is hence natural to define a generalized
Ginsburg criterion by the ratio (for D$<4$):
\bb \label{criterion} R_l=\ff{\int_0^l G(\bm{r})\dd
\bm{r}}{\int_0^l \langle m(\bm{r})\rangle ^2 \dd \bm{r}}\simeq A
\left( \ff{\xi}{l} \right) ^D. \ee
Here, $G(\bm{r})$ is the two-point correlation function which is
roughly equal to $1/r^{D-2+\eta}$ for $r<\xi$ and 0 for $r>\xi$.
$\langle m(\bm{r}) \rangle$ is the local magnetization, which
scales like $\xi^{-\beta/\nu}$. The second and approximate
equality is obtained using the scaling hypothesis and is true
independently of the value of the critical exponents.  The
traditional Ginsburg criterion is recovered for $l\sim\xi$ and it
is unsatisfied for all dimension D$<4$ ($A\simeq 1$). The main
result of ref.~\cite{CFH1} is that along the locus of temperatures
$\Tet$, while the correlation length diverges with $L$ as expected
for a critical phenomenon, its amplitude is small compared to $L$:
$\xi/L \simeq 0.03$. Therefore we have $R_L(T^*(L))\ll 1$ and  we
expect that the fluctuations at the integral scale $l=L$ could be
described, to an excellent approximation, by a perturbative
development. Such an approach cannot of course give correct
critical exponents but this is consistent with the observation
that such behaviour is, to a very good approximation, independent
of the exponents at hand~\cite{bramwell00,Banks05}. From the point
of view of renormalisation, we are expanding about the zero
temperature fixed point, which is Gaussian. The major contribution
to the fluctuations will come from length scales of the order of
the correlation length. From the microscopic scale up to $\xi$ the
effective action stays close to the non-trivial fixed point, only
to cross over towards the zero temperature fixed point for $l
\simeq L$. The calculation approximates the behaviour at
this scale with contributions estimated from the zero temperature sink. \\
To make the perturbative development we define the ``fast'',
fluctuating variables $\theta_i=\phi_i-\phi_0$ which we separate
from the ``slow" variable $\phi_0=\sum_i\phi_i/N$. Here the terms
``fast" and ``slow" are taken from the dynamical point of view as
we will see in detail below. We take the $\theta_i$ to be small.
The effective action, expanded to $lowest$ order in the $\theta_i$
and Fourier transformed can be separated into two parts:
\bb\label{res0}
\mathcal{Z}&\propto&
\int \dd \phi_0\prod_{\qv \in S_1}\dd \theta_{\qv} \dd \theta_{-\qv}
\exp -\action[\phi_0,\theta_{\qv}],\\
\nn \action[\phi_0,\theta_{\qv}]&=&\action_0(\phi_0)
+\sum_{\qv\in S_1}G_{\qv}^{-1}\theta_{\qv}\theta_{-\qv},\; G_{\qv}^{-1}=1-(1-\ttha^2 )K_{\qv},
\\ \nn
\action_0(\phi_0)&=&N\left[K(D+\lambda)\phi_0^2-\log\cosh(2K(D+\lambda)\phi_0+\beta
H)\right],
\ee
where $S_1$ is the minimal set of Fourier modes $\qv=(2\pi
p/L,2\pi q/L)\neq (0,0)$ that, in addition to the operation
$\qv=(2\pi(L- p)/L,2\pi(L-q)/L)$, fills up the entire Brillouin
zone except the zero mode. This prevents us to count twice the
quantities $\theta_{\pm\qv}$. 
We also define $\ttha=\tanh
[2K(D+\lambda)\phi_0+\beta H]$ and $K_{\qv}=2K(\lambda +\cos
q_x+\cos q_y)$. The total magnetization is then given by
$m_{\mathrm{tot}}\simeq \ttha-2\ttha(1-\ttha^2)\sum_{\qv\in
S_1}K_{\qv}\theta_{\qv}\theta_{-\qv}/N$.
\section{Dynamical approach}
To this lowest order, the action (\ref{res0}) looks similar to
that of the $2D$-XY model with a propagator that is a function of
$H$ and $\phi_0$. Setting $\phi_0$ constant would make it truly
XY-like, with a massive propagator. However, at zero field this is
not the case as we have here a finite size system. Hence, in
dealing with the effective action we have to take into account the
fact that there is no rigorous symmetry breaking and that the
equilibrium, low temperature magnetization is strictly zero, in
zero field. For this reason we choose a dynamical approach
separating time scales for fluctuations about a local free energy
minimum from those for a passage from one local minimum to the
other. This separation of scales defines the fast modes and slow
modes of evolution. At low temperature, or in finite field, the
slow, or ergodic time scale will be outside the numerical or
experimental observation time scale and we expect fluctuations
around a single minimum. As the critical point is approached, we
expect this separation of scales to be no longer possible with the
result that the symmetry is restored. \\
We can now follow the Langevin dynamics from the action
(\ref{res0}). Defining the fields $\phi_{\bm{q}}^{(1)}$ (resp.
$\phi_{\bm{q}}^{(2)}$) as $\mathrm{Re} \left(\theta_{\bm{q}}\right)$
(resp. $\mathrm{Im} \left( \theta_{\bm{q}}\right) $), we obtain the
following Langevin equations for the fast and slow degrees of
freedom:
\bb \dot{\phi}^{(\alpha)}_{\qv}(t)&=&-2G_{\qv}^{-1}
\phi_\qv^{(\alpha)}(t)+\eta^{(\alpha)}_\qv(t),\;
\dot{\phi}_0(t)=-\ff{\delta \action_0[\phi_0(t)]}{\delta
\phi_0(t)}+\eta_0(t), \label{eq_langevin} \ee
where the $\eta$ are Gaussian $\delta$-correlated noise:  $\langle
\eta^{(\alpha)}_{\qv}(t)\eta^{(\alpha')}_{\qv'}(t')\rangle=2
\delta_{\alpha,\alpha'}\delta_{\qv,\qv'}\delta(t-t')$, and
$\langle \eta_{0}(t)\eta_{0}(t')\rangle=2\delta(t-t')$.
%
%
%
\begin{figure}[t]
\begin{minipage}[]{\linewidth}
\begin{center}
\subfigure[\label{testa}Comparison at $T<\Tc$ and $H=0$]{
\includegraphics[width=0.45 \linewidth]{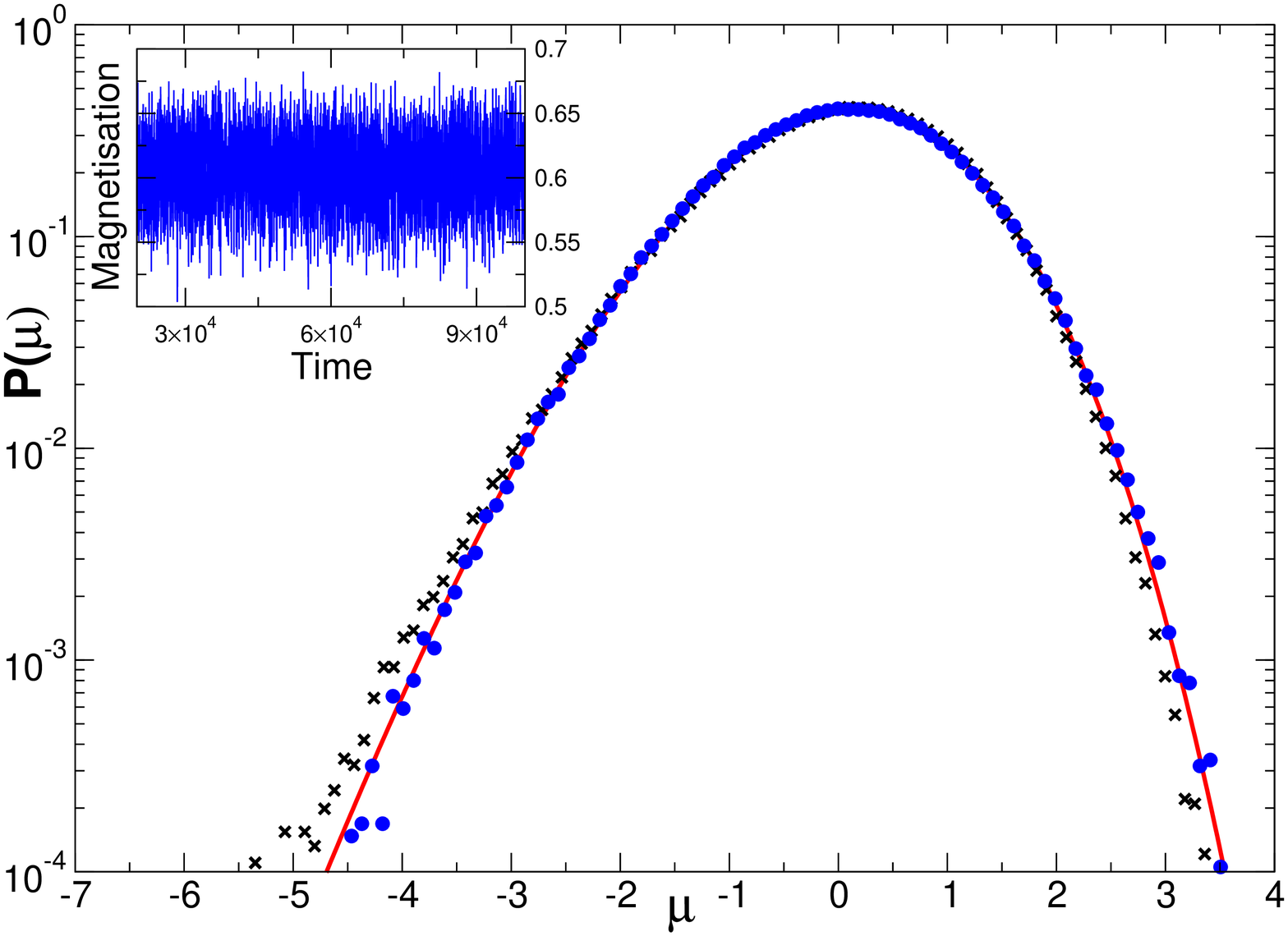}}
\subfigure[\label{testb}Comparison at $T=T^*$ and $H=0$]{
\includegraphics[width=0.45 \linewidth]{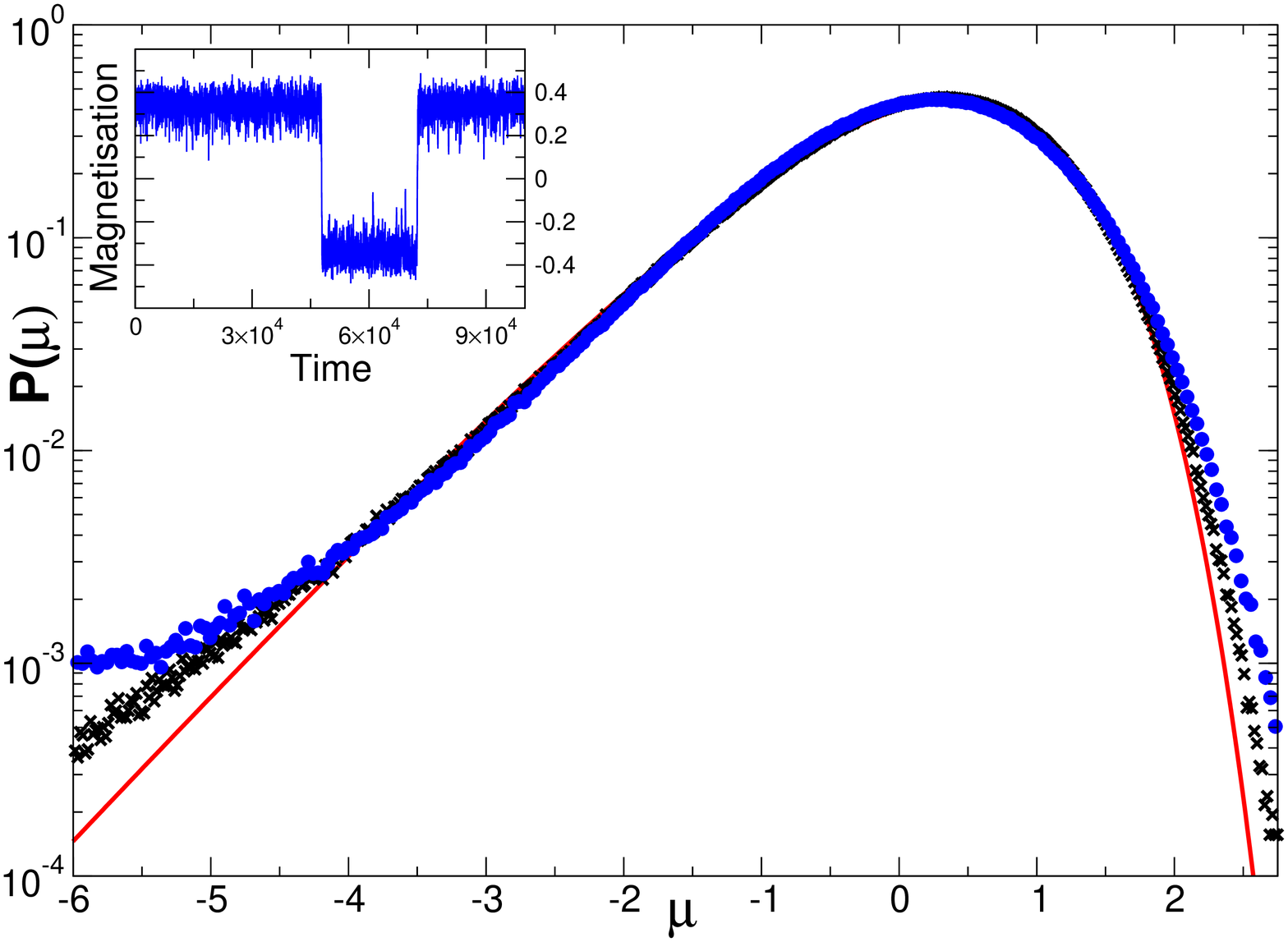}
} \caption{\label{test}Numerical test of the Langevin dynamics,
for $\lambda=2.5$ and $L=65$. (a) The data ($\times$) are Monte
Carlo simulations at $T=1.7$ and $L=64$.The data ($\bullet$) are
obtained by Langevin dynamics from equation ( \ref{eq_langevin})
at $T_\mathrm{dyn}=\Tc^{\mathrm{mf}}\times 0.74$. The plain curve
corresponds to the distribution of the Gaussian model ($L=64$)
with mass $M=3.5\times 10^{-1}$ \cite{portelli}. (b) The plain
curve is the BHP distribution \cite{bramwell01}. The data ($
\times$) are Monte Carlo simulations for $L=64$ system at
$T^*(L)=\Tc \times 0.94$ \cite{CFH1}. The data ($\bullet$) are
from the Langevin dynamics (\ref{eq_langevin}), at
$T^*(L,\lambda)=\Tc^{\mathrm{mf}}\times 0.95 $. Inset: examples of
magnetization dynamics.}
\end{center}
\end{minipage}
\end{figure}
%
%
%
%
%
We have now all the ingredients to compute $P(m,\tau)$, the PDF of
the instantaneous magnetization $m$ at time $\tau$. It is given
by, $ P(m,\tau)=\langle \delta\big(m-m_\mathrm{tot}(\tau)\big)
\rangle_{ \{\eta_0,\eta_{\qv}\} } \label{defPDF}.$ As
previously~\cite{LPF96,bramwell01} we introduce a reduced scaling
variable for the magnetization. Here we are interested in
fluctuations around the typical value of $m$, $\bar{m}=\langle
m_\mathrm{tot}\rangle_{\{\eta_{\qv }\}}$, where the average is
performed over all noise except $\eta_0$. Similarly we define the
width of the distribution as $\bar{\sigma}^2=\langle
m_\mathrm{tot}^2 \rangle_{\{\eta_{\qv }\}} -\bar{m}^2$, and the
reduced magnetization $\mu_\mathrm{tot}=(m-\bar{m})/\bar{\sigma}$.
We stress that $\bar{m}$ and $\bar{\sigma}^2$ are \emph{not} the
mean and variance of the instantaneous magnetization, as we
have not averaged over the noise $\eta_0$. \\
At this stage of our derivation, it is useful to numerically check
the approximations and assumptions made so far, \textit{i.e.} the
perturbation expansion, the Langevin dynamics and the logic of the
separation of time scales. In figure \ref{testa} we show data
generated by integrating numerically equations (\ref{eq_langevin})
for a temperature below the mean-field critical temperature, at
zero field. This shows that perturbation theory can indeed capture
the first departure from Gaussian fluctuations, as the critical
point is approached. In figure \ref{testb} we compare the PDF
obtained from Monte Carlo simulations at $T^*(L=64)=\Tc \times
0.94$ \cite{CFH1} with the results of Langevin dynamics at
$T^*(L=65,\lambda=2.5)=\Tc^{\mathrm{mf}}\times 0.95 $. As one can
see the agreement is good and the data give an equally good fit to
the BHP function. Differences appear in the wings of the
distribution. The perturbation scheme overestimates the value of
the PDF, illustrating the limit of its validity. However, it is
clear that, at least in the case where the symmetry is effectively
broken and where our criterion (\ref{criterion}) is satisfied, the
scheme captures the magnetic fluctuations of the Ising model to an
excellent approximation. Our scheme has to be compare with
mean-field treatment of Zheng~\cite{zheng03} where a simple
approximation or ansatz can also describe most of the features.
Here however, the connection with the XY-model is revealed through
a separation of the ``fast" variables, described by spin waves
like excitations, and ``slow" variables, representing the
evolution of the global magnetization within the effective
potential minima.
\section{Probability density function of the fluctuations}
Using the integral representation of the Dirac function,
$P(m,\tau)$ can be written as a path integral over the noise
$\eta_0$ and $\eta_\qv$. As the equations on $\phi_\qv^{(\alpha)}$
are linear (\ref{eq_langevin}), they can be integrated out if we
assume that $\phi_0$ is slowly varying with time. This is true at
low temperature: $\phi_0$ is the instantaneous magnetization
within the mean-field approximation, and the fluctuations around
this value are represented by the $\phi_\qv^{(\alpha)}$. The
amplitude of the fluctuations of the global variable $\phi_0$ in
equation (\ref{eq_langevin}) are scaled by the factor $1/N$ coming
from ${\cal S}_0$, and therefore $\phi_0$ does not, at least at
low temperature, venture far from the energy minima. In this case,
the quantity $G_{\qv}$, which depends on $\phi_0$, can be
considered as constant. That is, we can replace  the propagator by
its time averaged value during the period $\tau$.\\
After some algebra, defining $g_k=2\sum_{\qv \in S_1}\left[K_\qv G_\qv \coth
(\tau/G_\qv)\right]^k/N^k$, one finally obtains
\bb
P(m,\tau)=\int \D \eta_0(t) \exp
\left(-\ff{1}{4}\int _0^\tau \eta_0^2(u) \dd u \right)
\Pi_{XY}[\phi_0](\mu_\mathrm{tot},\tau), \label{lienXYising} \\
\nn
\label{Pi}
\Pi_{\mathrm{XY}}[\phi_0](\mu_\mathrm{tot},\tau)=\int_{-\infty}^{+\infty} \ff{\dd x}{2 \pi
\bar{\sigma}} \exp\left(ix
\mu_\mathrm{tot}+\fd\sum_{k=2}^{\infty}\ff{g_k}{k}\left (\sqrt{\frac{2}{g_2}}ix\right )^k
\right).
\ee
The function $\Pi_{\mathrm{XY}}$ is the PDF for the 2D-XY model in
the spin wave approximation with a massive propagator,
$G_\qv^{-1}\varpropto q^2+M(\phi_0)^2$ \cite{portelli}. Here
$\Pi_{\mathrm{XY}}$ depends on the temperature as the mass varies
with temperature through $\phi_0$. In the
limit $M\rightarrow0$, the temperature dependence disappears
and $\Pi_{\mathrm{XY}}$ becomes precisely the BHP function.\\
In order to obtain the PDF for $m_\mathrm{tot}$, we would have to
evaluate the last path integral over $\eta_0$. This integral is
related to the non linear Langevin equation (\ref{eq_langevin}).
At $T=0$ the dominant solution of the equation of motion is
$\phi_0^{(0)}(t)=\mathrm{constant}$, which is a solution of
$\delta {\cal S}_0/\delta \phi_0(u)=0$: the mode $\phi_0^{(0)}$
does not have any dynamics and it affects the PDF only by imposing
a finite mass $M(\phi_0^{(0)})$. Physically this means that the
PDF for the Ising model at low temperature should be the same as
that obtained for the 2D-XY model with a small magnetic field.
This is exactly what we have observed in figure \ref{testa}. At
finite temperature however non constant solutions exist: these are
the instantons associated with the non-linear Langevin equation
(\ref{eq_langevin}). Expressing the path integral in
(\ref{lienXYising}) over $\eta_0(t)$ as a path integral over
$\phi_0(t)$, using equation (\ref{eq_langevin}), it follows that
the time dependent solutions extremize the integral $\int
_0^{\tau}[\dot\phi_0(u)+\delta {\cal S}_0/\delta \phi_0(u)]^2 \dd
u$. This leads to
\be\label{integrChemin1}
P(m,\tau)&=&\int \D \phi_0(t)
\exp
\left [
-\ff{1}{2}\big({\cal S}_0[\phi_0(\tau)]-{\cal S}_0[\phi_0(0)]\big)\right ]\times
\\ \nn
& &
\exp
\left[-\ff{1}{4}\int_0^{\tau}\left (
\dot\phi_0^2(u)+
\left (
\frac{\delta {\cal S}_0}{\delta \phi_0(u)}
\right )^2-2
\frac{\delta^2 {\cal S}_0}{\delta \phi_0(u)^2}
\right )
\dd u
\right ]
\Pi_{\mathrm{XY}}[\phi_0](\mu_\mathrm{tot},\tau).
\ee
The second order derivative of ${\cal S}_0$, which comes from the
Jacobian of the transformation, is of order $N$ compared with
order $N^2$ for the other terms and is therefore negligible. To simplify
the analysis, we assume periodic boundary conditions, $\phi_0(0)=\phi_0(\tau)$,
and the first argument of the
exponential in (\ref{integrChemin1}) vanishes. The non constant
solutions $\phi_0^{(k)}(u)$, $k>1$ verify
$\dot\phi_0(u)^2/2=V_0-V_{{\rm eff}}[\phi_0(u)]$, where
$V_0>0$ is a constant of motion that depends on the instanton
trajectory and $2V_{{\rm eff}}[\phi_0(u)]=-[\delta {\cal
S}_0/\delta \phi_0(u)]^2$ is an effective inverted potential. From
(\ref{integrChemin1}) we define an effective action $S_{{\rm
eff}}$ whose $\phi_0^{(k)}$ are the classical time dependent
solutions: $S_{{\rm eff}}[\phi_0(u)]=\int_0^{\tau}\left (
\fd\dot\phi_0^2(u)-V_{{\rm eff}}[\phi_0(u)] \right ) \dd u$.
Expressing $\dd u$ as function of $\dd \phi_0$, the constant $V_0$
satisfies the equation $\int \pm\frac{d\phi_0}{\sqrt{2(V_0-V_{{\rm
eff}})}}=\tau.$ The sign is positive (resp. negative) when $\phi_0$
is increasing (resp. decreasing) with time.  Generally, we
assume that $V_0$ tends exponentially to zero when $\tau$ is
large. For an action with two wells ${\cal
S}_0=-a\phi_0^2/2+\phi_0^4/4$, with $a\propto T_c-T$, we can
evaluate this constant. Indeed, if we consider a trajectory going
from $\phi_0(0)=\sqrt{a}$ to $\phi_0=0$ at some later fixed time,
then going back to $\sqrt{a}$ at time $\tau$, this corresponds to
solving the following equation:
$2\int_{0}^{\sqrt{a}}\frac{d\phi_0}{\sqrt{2(V_0-V_{{\rm
eff}})}}=\tau$. When $\tau$ is large, $V_0$ has to be small, and
the main contributions from the previous integral come from the
end points: $V_{{\rm eff}}\simeq -2a^2(\phi_0\mp\sqrt{a})^2$ for
$\phi_0$ close to $\pm\sqrt{a}$, and $V_{{\rm eff}}\simeq
-a^2\phi_0^2/2$ for $\phi_0$ close to 0. It is then easy to show
that $V_0\propto \exp(-\sqrt{2}a\tau)$ and it is hence legitimate
to discard this constant for large time $\tau$. The effective
action $S_{\mathrm{eff}}$ can be simplified in this case, and
is equal to $\Delta {\cal S}_0^{(k)}$, the sum of the energy
barriers crossed by the instanton $\phi_0^{(k)}$ during the time
$\tau$. Finally, the distribution (\ref{integrChemin1}) can be put
into the following form
\be\label{integrChemin2}
P(m,\tau)\sim \Pi_{\mathrm{XY}}[\phi_0^{(0)}](\mu_\mathrm{tot})+
\sum_{k\ge 1}
C_k\exp
\left (
 -\Delta {\cal S}_0^{(k)}
\right )
\Pi_{\mathrm{XY}}[\phi_0^{(k)}](\mu_\mathrm{tot},\tau),
\ee
where the $C_k$'s are coefficients related to the Gaussian
fluctuations around the saddle point solutions $\phi_0^{(k)}$.
These instantons restor the system symmetry. Below the critical
region they appear only on exponentially large time scales, and
can be neglected on the time scale of any observations. Near the
critical point they appear more frequently and lead to a phase
transition in the finite system. It is extremely challenging to
exactly compute the contribution from the instantons~\cite{alex93}
and to do so would, in any case, give only an approximate
description of the true critical dynamics, so we do not attempt it
here. Rather, the final expression (\ref{integrChemin2}) is
sufficient for our purposes as it exposes both the origin of
generic critical fluctuations and of the dependence on
universality classes through the generation of instanton like
excitations.
\section{Interpretation and generalization}
We have shown that at this level there are two distinct
contributions to the PDF for magnetic fluctuations in equation
(\ref{integrChemin2}). The first is a Gaussian action coming
directly from the perturbation expansion and as long as quadratic
fluctuations are present, such a term must appear. The limiting
case of zero mass corresponds BHP distribution. The Ising model
universality class does not appear in this term and in this sense
it is superuniversal. The fluctuations described here are
localized around the typical value of the magnetization and are
sensitive to the local geometry around this point in phase space,
not to the global structure of the phase space related to the
universality class. The second term comes from the contribution of
instantons that restore phase space symmetry and so is strongly
dependent on the universality class. It is the analogue of the
corrections computed exactly for the 2D-XY model within
the spin wave approximation \cite{Banks05,Mack05}\\
The existence of a point, or points, in the phase diagram where the
PDF of the 2D Ising model is close to the BHP distribution
\cite{CFH1} results from a compromise between these two terms: to be
as close as possible to BHP, we have to reduce the mass in the
Gaussian propagators of the first term. This is exactly what
happens as one approaches $\Tc$ and the instantaneous value of
$\phi_0$ reduces. At the same time the contribution from
instantons has to be small, which is not the case at $\Tc$:
reducing $\phi_0$ increases the probability of inducing an
instanton, taking the system from one local minimum to the other.
This occurs just as the correlation length becomes of the order of
the system size, hence the instanton contribution being small
corresponds exactly to our criterion (\ref{criterion}) being
satisfied. The temperature $T^*$ studied in \cite{CFH1} is the
point of best compromise. The application of a small magnetic
field breaks the symmetry, increasing the barriers $\Delta {\cal
S}_0^{(k)}$. The contribution from instantons is then reduced and
one can expect better agreement with BHP than in zero field. This
is just what is observed. However, while the agreement with BHP
can be excellent it is approximate and there will always be
corrections to it.

What other correlated systems show similar behaviour ? The
previous analysis suggests that generic behaviour reminiscent of
the D-dimensional Gaussian model should indeed be commonly
observed in equilibrium correlated systems in D-dimensions. We
show in fact that fluctuations of this generic type correspond to
perturbative corrections to the central limit theorem and that in
this sense such systems should be thought of as being weakly,
rather than strongly correlated. It suggests also that one should
expect variations. These variations can be large, coming from
non-linear instanton-like objects, taking the distribution far
from the generic asymmetric form shown in
Fig.\ref{testb}~\cite{CFH1,mal04}, or they can be smaller, coming
from the harmonic contribution itself. That is, the BHP function
is not a miraculous unique function for all circumstances, rather
distributions of the form of the first term in equation
(\ref{integrChemin2}) depend weakly on the boundary
conditions~\cite{bramwell01,racz02,rosso03}, on the
mass~\cite{portelli} and even on the temperature
\cite{Banks05,Mack05}.  Of these parameters the strongest
dependence is on dimension: we expect that 3D correlated
equilibrium systems are related to the 3D Gaussian model, which
has weakly asymmetric order parameter fluctuations and is not
critical~\cite{bramwell01,chamon04}. Interestingly, out of
equilibrium three dimensional systems show results resembling the
2D-XY model\cite{BHP,bramwell00,gleeson}, or the closely related
Gumbel distribution~\cite{racz02} characteristic of a
one-dimensional system with long range interactions not those of
the 3D-XY model. These results seem to suggest the presence of a
dimensional reduction in such systems that remains to be
explained. \acknowledgments It is a pleasure to thank S.T.
Bramwell, S.T. Banks, F. Delduc and P. Pujol for useful
discussions and comments.

\end{document}